# Estimation of the reconstruction parameters for Atom Probe Tomography

B. Gault[1], F. de Geuser[2], L.T. Stephenson[1], M.P. Moody[1], B.C. Muddle[2] and S.P. Ringer[1]

1 Australian Key Centre for Microscopy & Microanalysis, The University of Sydney, NSW, 2006, Australia.

2 ARC Centre of Excellence for Design in Light Metals, Department of Materials Engineering, Monash University, Clayton, Victoria 3800, Australia

## Abstract

The application of wide field of view detection systems to atom probe experiment emphasizes the importance of careful parameter selection in the tomographic reconstruction of the analysed volume, as the sensitivity to errors rises steeply with increase in analysis dimensions. In this paper, a self-consistent method is presented for the systematic determination of the main reconstruction parameters. In the proposed approach the compression factor and the field factor are determined using geometrical projections from the desorption images. A 3D Fourier transform is then applied to a series of reconstructions and, comparing to the known material crystallography, the efficiency of the detector is estimated. The final results demonstrate significant improvement in the accuracy of the reconstructed volumes.

**Introduction**

Over the last 10 years, the field-of-view in the atom probe tomography (APT) technique has been significantly expanded (Seidman, 2007; Kelly and Miller, 2007). In particular the recent implementation of the Local Electrode Atom Probe (LEAP$^{TM}$) and the use of laser pulsing has enabled a reduction in the flight path without affecting the elemental sensitivity, the mass resolving power or the spatial resolution (Kellogg & Tsong, 1980; Tsong et al., 1982; Kelly et al., 1996; Kelly et al., 2004; Gault et al., 2006). Figure 1 is an example of APT data from a AlAgCu alloy, heat treated at 150°C for 120 minutes, presented in the form of a 3D atom map. The high spatial resolution of the technique is readily apparent: atomic resolution is achieved in at least one dimension, one of the main strengths of APT in comparison to other nanoscale characterisation techniques. Essentially, an atom probe can be considered as a point projection microscope that enables the detection of single atoms removed from the surface of the specimen via the application of a very intense electric field (Müller et al., 1968). Coupled to a time-of-flight mass spectrometer and using a position sensitive detector (PSD), the atom probe becomes a nano-analytical microscope capable of mapping the distribution of the atoms in three dimensions (Panitz, 1973; Panitz, 1974; Panitz, 1978; Cerezo et al., 1988; Blavette et al., 1993).

The capacity for quantitative observation of 3D chemical distribution chemistry at the atomic level has makes APT a very powerful technique for physical metallurgy, enabling new insights into the mechanism of phase formations, precipitation, interface science or ultrafine scale materials characterisation (Hono et al., 1992, Miller et al. 1996; Edwards et al., 1998; Blavette et al., 1999; de Geuser et al., 2006; Moody et al., 2007; Timokhina et al., 2007; Seidman, 2007). Further, the recent implementations of laser pulsing is now facilitating a broader impact on materials science research, e.g.

in the field of semiconductors science and technology (Thomson et al., 2007), as recently reviewed by Kelly et al. (2007).

The accurate reconstruction of the atomic structure and chemistry in materials requires knowledge of how the atoms are projected from the specimen onto the detector. Although the model typically used to reconstruct APT volumes was developed for atom probes with a physical angular aperture of about 10 degrees (Blavette et al., 1993; Bas et al., 1995), the point-projection model appears to remain valid for the more recent commercial atom probes that can have angular aperture of more than 40 degrees. However, due to the subsequent increase in the volume of the specimens that can now be analysed, it becomes increasingly difficult to fine-tune the reconstructions in 3D such that, for example, atomic planes remain undistorted and precipitate shapes physically meaningful.

In this model, there are three main reconstruction parameters: the image compression factor, the field factor and the efficiency of the PSD. The latter is clearly a parameter linked to the instrument. The first two are more closely linked to the morphology and nature of the specimen itself (Hyde et al., 1994), hence the recent interest in determination of specimen geometry (Gorman, 2007). In this study, an original methodology is developed and applied to obtain accurate values for these reconstruction parameters, validating the use of the current reconstruction model to wider field–of-view instruments, at least in the case of metals and metallic alloys.

# Fundamentals of Reconstruction in APT

The technique of atom probe is based upon the physical principle of field evaporation, a combination of ionization and subsequent desorption of surface atoms induced by an electric field. This is made possible by the application of an electric field, lowering the potential energy barrier binding atoms to the surface such that the thermal agitation of the surface atoms is sufficient to overcome this barrier. The probability of realizing field evaporation is thus critically dependant on the amplitude of the electric field.. This *evaporation field,* $F_e$, the theoretical electric field at which the height of the potential barrier is decreased to zero, is in the range of 10 to 60 V.nm$^{-1}$ for most metals ( Brandon, 1961; Müller, 1965; Tsong, 1971). To achieve such an intense electric field, the specimen is fabricated into the shape of a sharp needle, with a radius of curvature in the range of 50 to100 nm at the apex. The radius of curvature is assumed to be the radius of a sphere tangential to the surface at the apex of the tip. This specimen geometry enables evaporative fields to be attained for many materials even when the sample is subject to an applied voltage of just a few kV. At the apex of this specimen, the electric field F can be expressed as:

$$F = \frac{V}{k_f R} \quad (1)$$

where V is the applied voltage, R the radius of curvature and $k_f$ is a geometric field factor that accounts for the shape of the tip. The tip can be regarded as a hemispherical cap atop a truncated cone . The angle of the cone is called the shank angle of the tip. The concentration of the electric field at the apex of the tip depends on this angle and on its electrostatic environment and typically ranges between values of 2 and 8 (Gomer, 1961; Sakurai, 1973; Sakurai, 1977). Time controlled field evaporation of atoms is ensured by superimposing high-voltage or laser pulses upon the DC standing field applied to a tip. Further, the specimen is maintained under ultra-high vacuum (UHV) conditions, and its temperature is carefully controlled, depending on the material, to minimise thermal surface diffusion of atoms prior to evaporation. Generally, it is cooled to cryogenic temperatures, down to 15 K. Figure 2 shows the configuration of a modern commercial atom probe.

The electric field is highly divergent at the apex of the tip, due to its curvature. The emitted ions follow the electric field lines, which gives rise to a projection with a very high magnification. Recording the successive arrival positions of the atoms on the detector thus results in a highly magnified image of the surface. Further, since the surface of the specimen can be regarded as the intersection between the crystal lattice and a hemisphere, the image of the surface will reveal a polar structure, characteristic of the crystallography of the material. This principle has been utilised in both the field ion microscope (FIM) and in the field desorption microscope (FDM) to study crystallographic structures, defects and grain boundaries (Müller et al., 1956 ; Panitz, 1973; Waugh, 1977; Tsong, 1990). The two-dimensional image formed by the arrival of ions on the detector during an APT analysis is a desorption pattern that will also sometimes reveals a pole structure (Fig. 3). Provided that these pole structures are well resolved and the field of view sufficiently large, crystallographic directions can usually be identified.

In APT, the use of a position-sensitive detector (PSD) enables the collection the detected positions of the atoms on the detector $(X_D, Y_D)$. This detector is based on an assembly of micro-channel plates (MCP) and a delay-line detector. An inverse point-projection algorithm is used to calculate the original atomic positions (x, y, z) at the surface of the specimen. The geometric configuration of the projection is shown on Fig. 4. The image magnification can be expressed as :

$$M = \frac{L}{\xi R} \qquad (2)$$

with R the radius of curvature and L the flight length. The compression factor, $\xi$, accounts for the modification of the field lines due to the shape of the tip. All the field lines are assumed to cross at a single point (P) on the tip axis and $\xi$ can vary between 1 (radial projection) and 2 (stereographic projection). In many studies ξ has been shown to lie somewhere near the middle of these two values (Brandon, 1964; Newman et al., 1967).

The x and y coordinates can therefore be deduced as:

$$x = \frac{X_D}{M} \quad \text{and} \quad y = \frac{Y_D}{M} \tag{3}$$

As the tip usually has a non-zero shank angle, the radius of curvature increases during the analysis. A possible way to deduce the evolution of the radius of curvature is by using the total voltage applied to the specimen. Assuming that the electric field at the apex of the tip is equal to the evaporation field, $F_e$, and constant across the imaged surface of the tip and that $k_f$ is constant, then the radius of curvature can be derived by rearranging Eq.1:

$$R = \frac{V}{k_f F_e} \tag{4}$$

The electric field is screened on the Debye length, much smaller than the size of a single atom in the case of good conducting materials, and therefore only atoms at the very surface of the specimen are likely to be field evaporated (Suchorski et al., 1995). Due to this property, the sequence of evaporation of the ions can be used to deduce the in-depth dimension (z). To ensure that the total detected volume is consistently correct, an increment in depth is calculated for each detected atom. This increment corresponds to the atomic volume distributed on the whole surface of analysis:

$$dz = \frac{\Omega}{\eta S_a} \tag{5}$$

where $\Omega$ is the atomic volume, $\eta$ the detection efficiency and $S_a$, the total analysed surface, given by:

$$S_a = \frac{S_d}{M^2} \tag{6}$$

with $S_d$ the surface area of the PSD, and M the magnification. Assuming a constant electric field equal to the evaporation field, an expression for dz can be derived by combining Eq. 2, 4, 5 and 6:

$$dz = \frac{\Omega L^2 k_f^2 F_e^2}{\eta S_d \xi^2 V^2} \qquad (7)$$

The z position is calculated by sequentially cumulating this increment, *dz*, and applying a corrective term dz' to take into account the curvature of the specimen for every individual atom:

$$dz' = R\left(1 - \sqrt{1 - \frac{x^2 + y^2}{R^2}}\right) \qquad (8)$$

A three-dimensional map of the positions of the evaporated atoms can be progressively constructed. The effect of angular tilt of the specimen with respect to the direction normal to the plane of the detector may also be accounted for in the 3D reconstruction. This effect can be significant in instruments where gimbals are needed to orient the specimen so as to bring a selected feature into the field-of-view of the detector (Bas, 1995). However, these corrections have little effect at low solid angles and are not generally required in wide field-of-view detector systems such as the LEAP™, where the long axis of the specimen is fixed, parallel to the normal to the detector plane.

In fact, the reconstruction parameters, η, ξ, $F_e$ and $k_f$, are generally both instrument and specimen dependent. In principle, these quantities can be determined by a careful combination of FIM and APT (Hyde, 1994). However, this approach requires great care and is time consuming. Further it is usually not possible due to image contrast issues in many materials. It is more usual to assume that η and ξ depend only on the instrument and can be determined in occasional instrument calibration experiments. Conversely, there is limited information available for $F_e$ values and so known values for materials that are similar in composition and phase constitution to the specimen of interest are commonly applied.

It has previously been shown that APT spatial resolution in depth can be better than one interplanar spacing in a given {hkl} direction and better than 1 nm in the x-y plane (Vurpillot et al., 2001). This resolution is sufficient to resolve atomic planes of planes near normal to the direction of the tip axis. If the planes can be correctly identified, the inter-planar spacing can be used to more accurately reconstruct the volume. The resultant inter-planar spacing observed after the reconstruction can be refined or fitted by adjusting the above reconstruction parameters. However, it is significant that these parameters have a precise physical meaning and so far as is possible, these terms ought not be fitted arbitrarily. These terms are not independent from one another and, significantly, they can be expected to change between instruments, specimens and from one particular experiment to another on a given tip. Thus, to generate the most accurate reconstruction of the whole volume a new approach is now described.

# Experimental

Pure aluminium specimens of very high purity (> 99.999%) metal were prepared using standard electropolishing with 33% nitric acid in methanol under a binocular microscope. They were analysed using an Imago LEAP-4 equipped with a 4 cm delay line detector at a pulse fraction of 0.2. The specimen was maintained below 20 K under ultrahigh vacuum conditions (< 4.5 $10^{-9}$ Pa). Initial tomographic reconstructions were performed using the standard methods described above.

# Determination of the reconstruction parameters

*Compression factor ξ*

As described in the previous section, the compression factor arises from the deformation of the field lines around the apex of the tip. Although areas of lower density of ions striking the detector have previously been observed with earlier instruments and attributed to crystallographic poles and zone axes (Waugh et al., 1976), the field-of-view was too small to enable identification of these features. However, the wider field-of-view now available makes it possible to index the pole structure in the desorption image on the detector, particularly in the case of metals and dilute alloys, as shown for pure aluminium on Fig. 3. Once the poles have been identified, the compression factor can be expressed simply as the ratio of the angle between two crystallographic directions $\theta_{crys}$ and the actual angle $\theta_{obs}$ observed on the desorption pattern obtained during the analysis (Fig. 4.). The observed angle is simply calculated measuring the distance, D, between the centre of two given poles:

$$\theta_{obs} = \arctan\left(\frac{D}{L}\right) \qquad (9)$$

In this way, the compression factor was determined both by measuring the distances between the centres of all the poles identified in Fig.3 on a desorption image containing $5 \times 10^5$ atoms and plotted after 10 million atoms were collected. The average value from all poles was $\xi = 1.70 \pm 3\%$. This compares closely with the value of $\xi = 1.71 \pm 3\%$, which was obtained using only the three most prominent poles in the image, corresponding to the {001}, {113} and {102} plane families. They will respectively be denoted 002, 113 and 042. The very good agreement between these values enabled the use of these three poles to estimate the compression factor at different depths during the analysis. The distance between the centres of these poles was measured on desorption maps containing $5 \times 10^5$ atoms. The corresponding angles were determined and the compression factors have been plotted as a function of the number of atoms in regard to the evolution of the potential in Figure 5. Two distinct behaviours can be observed: an initial region of the graph, corresponding to the very first atomic layers evaporated, during which the compression factor varies significantly, and a plateau region corresponding to the remainder of the acquisition. At the very beginning of an APT experiment there is a finite period in which the specimen is still developing its temperature dependant steady-state shape, called equilibrium shape in the literature (Vurpillot et al., 1999). This period accounts for the initial fluctuations in the measured compression factor value. Significantly, however, the compression factor is constant for most of the experiment. An average value of $1.70 \pm 3\%$ was determined along the plateau, which is the value that will be used for the reconstruction.

### *Radius of curvature R and Field Factor $k_f$*

The reconstruction model requires an instantaneous value of the radius of curvature for each evaporated atom. Equation 4 indicates that this radius of curvature can be deduced from the applied voltage, provided that the value of $F_e k_f$ is known. Interestingly, only the product of $F_e$ and $k_f$, rather than their independent values, is necessary. Methods to estimate the radius of curvature from FIM micrographs have been previously developed (Dreschler et al., 1958; Miller et al., 1996). As shown by Waugh et al. the desorption figure reveals patterns similar to the FIM image, even if they originate from different mechanisms: the desorbed ions come from the surface itself whereas imaging ions come from the ionization zone at a few Angstrom above the surface (Waugh et al., 1976). The desorption pattern can be used to measure the specimen radius of curvature.

To carry out these measurements, a 2D map containing only a small number of atoms, corresponding to less than one atomic layer, has been plotted (Fig.6(a)). On this map, rings, characteristic of pole structure, can be observed, for example, around the 002 pole. The rings correspond to the successive terraces projected on the flat surface of the detector as shown in the Fig. 6(b).

The sizes of the terraces depend on the radius of curvature and on the depth of the first terrace, which is in effect the distance between the first terrace and the hemisphere used to model the tip surface shape. In depth, two {hkl} terraces are separated by a known distance $d_{hkl}$, which has a value of 0.2025 nm for a 002 pole in pure aluminium. Following the approach outlined by Drechsler and Wolf (1958), a linear relationship between the angle and the depth increment can be established based only on geometrical considerations:

$$1 - \cos\theta_n = \frac{d_{hkl}}{R_0} n + \frac{a_0 - d_{hkl}}{R_0} \qquad (10)$$

where $a_0$ is the depth of the first terrace with respect to the hemispherical cap of radius $R_0$ and can be defined as the remainder of the division of $R_0$ by $d_{hkl}$. The term *n* is the number of the terrace. Note that *n* is a natural number greater than 1, indeed, as $0 < a_0 < d_{hkl}$, eq..10 is only valid for values n > 0. The diameters of successive rings can be measured directly from the 2D detector map and the corresponding angles $\theta_n$ determined by linear scaling of distances on the detector map to known angles from the poles. Plotting the quantity $1-\cos\theta_n$ as a function of the terrace index, the slope of the corresponding straight line provides a measure for the radius of curvature. The depth of the first terrace can subsequently be deduced from the intersection with the y axis.

In practical terms, to generate the plot shown in Fig.7, the tip was first allowed to develop into an equilibrium shape after collecting ~10 million ions. Determinations of $1-\cos\theta_n$ as a function of the index number, n, of the terraces (Eq. 9) were fitted to a straight line where the slope is inversely proportional to the radius of curvature, R. This method was applied to estimate the local radius of curvature around the three major poles, 002, 113 and 204. $R_0$ was determined by taking the average of the estimated radii of curvature on each pole was measured to be 47.15 nm ± 6%.

This procedure was repeated to measure the radius of curvature at several stages through the experiment for different numbers of detected ions. The values of the radii and voltages are reported in Table 1 and plotted in Fig. 8. According to Eq. 1, the radius of curvature of the tip is directly proportional to the voltage and so a straight line has been fitted constrained to go through 0. The value generally used for the field necessary to evaporate an $Al^+$ ion is $F_e=19 Vnm^{-1}$ (Tsong, 1978), an average value of $k_f= 3.60 \pm 6\%$ was determined assuming this value.

*Detector Efficiency η*

As discussed above, accurate tomographic reconstruction requires determination of the depth in z-direction and it is here that an accurate estimate of the detector efficiency, η, is critical. An approach is proposed, that utilises an initial 3D reconstruction of the volume using the values determined above for the reconstruction parameters ξ, $F_e$ and $k_f$ and using an arbitrary value for detection efficiency (e.g. η

=0.5). This initial tomographic reconstructed atom map was then carefully examined. The value of the detection efficiency η was adjusted until the distance between successive {002} planes, equal to $d_{002}$, was equivalent to the known value This, in effect, fixes the ratio $(k_f F_e)^2/\eta$.

However, it is important to take into account that a significant uncertainty in the estimation of $k_f F_e$ is inherent in the radius of curvature by the method described above. This uncertaintiy, which relates to the lack of precise information for $F_e$ and the sensitivity of $k_f$ to issues such as an evaporated ion's original position on the surface of the specimen, can detrimentally affect the precision of the value of the detection efficiency. To overcome this uncertainty another original method for determining the efficiency of the detector has been developed.

The approach requires that the value of $(k_f F_e)^2/\eta$ be maintained as constant as possible to ensure a correct spacing between the 002 planes across a series of separate tomographic reconstructions applying varying values η. The reconstruction was obtained using the entire surface of the MCPs, including the low efficiency region that can be observed close to the 113 pole in the Fig.3.. As the total depth of the reconstruction is kept constant, only the apparent analysed surface will be progressively increased, changing the apparent angle between the different crystallographic directions inside the reconstruction. This effect was assessed for each reconstruction using a Fourier transform (Camus et al., 1995; Vurpillot et al., 2001) from an equivalent small volume in every reconstruction, as depicted in Fig. 9(a) and (b). Figure 9(c) reveals more detail of the structure inherent within the reconstruction and highlights the Al atoms that constitute the {002} lattice planes, the spacing of which is used as a reference for this reconstruction approach. The transform around these (x, y, z) real space coordinates results in a diffraction pattern provided in Fig. 9(d). The indexed red spots are values which can be distinguished in the Fourier Transform of the APT data, whereas the un-circled spots were not detected.

Next, for the diffraction pattern obtained from each reconstruction, the apparent angle $\theta_{<001>-<113>}$ between the relevant <001> and <113> directions was measured. This angle was plotted as a function of the efficiency, Fig. 9. The correct detector efficiency can then be deduced from the value that corresponds to the known value $\theta_{<001>-<113>}$ = 25.24 degrees. Inspection of Fig. 10 shows that in this particular case, this approach led to an estimate of the detection efficiency of $\eta$ = 65%±7%. The corresponding $k_f F_e$ parameter was be re-evaluated using this value of $\eta$, giving a value of $k_f$=3.765. On this same basis, a corrected radius of curvature, R, was determined from Eq. 4 of about 48 nm, after 10 million atoms. This is compatible with the values obtained in the preceding section.

## Discussion

The proposed methodology involves the determination of the tomographic reconstruction parameters utilising information within desorption images available from the detector during an atom probe analysis of pure aluminium and is particularly amenable to wide field-of-view detection systems. There are effectively two steps to the approach. In the first step, the compression factor $\xi$ (Eq.2) was determined using desorption images. The small variations observed for the first few million evaporated atoms may be explained by the progressive changes that occur in tip shape as the field evaporation end form is obtained (Vurpillot, 1999). The compression factor rapidly reaches a consistent value after the collection of 2.5 x $10^6$ atoms with an average of $\xi$ = 1.703 thereafter. This indicates the onset of formation of an equilibrium shape for field evaporation. Although the compression factor slightly decreases, the variation never exceeds 5% and so the average value in the plateau region of the graph may be as regarded as effectively constant throughout the experiment, albeit averaged across the whole area of the detector.

The second step involves determination of the field factor, $k_f$. The approach here of fitting a straight line to the radius of curvature and the applied voltage (Fig.7) depends on an assumed value of the electric field for evaporation. Whilst the voltage was adjusted during the atom probe analysis so as to maintain a constant detection rate of atoms, the radius of curvature of the tip will progressively increase and so the total analysed area $S_a$ (Eq. 6) also increases. The detection rate is proportional to the factor $N_{at} \times P_{evap}$, where $N_{at}$ is the number of atoms that are susceptible to be field evaporated and $P_{evap}$ the probability of field evaporation. As the analysed area increases, consequentially $N_{at}$ also progressively increases. Thus, to maintain a constant rate of detection, $P_{evap}$ decrease proportionally. Its value, being directly proportional to the electric field at a given temperature, means that the electric field at the surface has to decrease progressively to keep constant the value of $N_{at} \times P_{evap}$. This way of controlling the experiment implies that the radius of curvature should not be strictly proportional to the voltage. A basic correction to the field factor has been applied to account for the variation of $S_a$ in the determination of the values of $k_f$ in each case. The corrected values have been reported in Table 1. The decrease in field factor during the analysis is attributed to the evolution of the tip shape and the average value was 3.77± 6%, which is similar to the value deduced from the Fourier transform calculations.

Strictly speaking, the values of both the image compression and the field factors are dependent on a combination of many different parameters, such as the shape of the specimen or possibly the electrostatic environment. Cerezo et al. (1999) have shown that the projection is actually controlled by the specimen itself and is not sensitive to the presence of the counter electrode located a few millimetres from the tip apex, but this assumption remains to be confirmed in the case of a local-electrode. However, both parameters are certainly not independent from each other. The measurements in this study clearly show their values progressively decreasing during the experiment. Significantly, these changes are relatively small effectively constant average values are available from this approach, where the depth of analysis is less than ~100 nm.

The radius of curvature, the field factor of the specimen and the the detection efficiency have all been determined using an iterative method based on successive Fourier transform calculations to calibrate the angle between two identified crystallographic directions. A value of 65% has been determined for the efficiency, knowing only the image compression factor that has been completely independently measured. This value is near the higher limit of the MCP efficiency, corresponding to the open area of its surface, but is in good agreement with the value recently found by Geiser et al. (2007) on the same detection system. The values of the field factor and radius of curvature that can be determined using this method are in very good agreement with the values obtained by a direct measurement on the desorption images.

As indicated by Eq.7, the in-depth increment is proportional to $\frac{k_f^2 F_e^2}{\eta \xi^2}$. This emphasizes the importance of independently determining the value for the image compression factor. Further it is quite straightforward to estimate an accurate value for this parameter. Indeed, even if it is still possible to adjust all the parameters based only on generating the right interspacing between the in-depth planes, this approach can be at the expense of introducing curvature effects into the planes.

It is significant to note that, although the actual projection parameter may vary slightly during the analysis as the shape of the tip evolves, a single set of reconstruction parameters leads to satisfactory reconstructions even on data sets up to 30 millions of atoms representing about 250nm in depth. Indeed, measurements of the $d_{002}$ interplanar spacing were shown to maintain a consistent value of around 2.01Å with variations no greater than 6%. Further, the diffraction pattern in Fig.11 shows very high index spots, up to {3, 3, 11}, corresponding to interplanar distances of less that 0.035nm. A large number of atomic planes directions are resolved, even with angular differences as high as 35 degrees between 224 and 002.

## Conclusion

In conclusion, conversely to other methods (Hyde et al., 1994), the procedure outlined in this study, utilises measurements solely from the atom probe experiment and does not rely on the assumption that the projection parameters are equivalent in FIM and atom probe configuration. This assumption is subject to question, as it is well known that both the presence of the image gas, the use of a pulsed electric field or the electrostatic environment can change the behaviour of the specimen and the projection of the ions. Additionally, the projection parameters can still be determined a posteriori on any suitable data set obtained on an atom probe to get better reconstructions, without the need for assumed values for the different parameters. This study also highlights that, despite its relative simplicity, the classical reconstruction algorithm for atom probe data set still gives very good results, even on wider field of view instruments. Atomic planes separated by less than 0.04 nm can indeed be imaged. Further, the new generation of atom probes have an angle of aperture nearly twice bigger, therefore, more information can potentilaly be obtained from the desorption pattern. However, the risk of curvature artefact is increased near the edges of the analysis, even if the reconstruction should be reliable on at least half of the analysed area.

## Acknowledgements

The authors are grateful for funding support from the Australian Research Council, which partly sponsored this work. The authors are grateful for scientific and technical input and support from the Australian Microscopy & Microanalysis Research Facility (AMMRF) at The University of Sydney.

# Figure captions

Table.1: Evolution of the radius of curvature, the voltage and the field factor as a function of the number of detected atoms.

Fig.1: APT 3D map of an Al-5.6Ag-0.84Cu (at. %) alloy. Only Ag atoms are represented (25%). GP regions are clearly observable (Note standard reconstruction parameters have been used for this reconstruction).

Fig.2: Experimental setup. UHV stands for Ultra-High Vacuum, PSD for Position Sensitive Detector, HV for High Voltage.

Fig.3: Desorption image obtained with the first 300000 atoms detected during a pure Al analysis (2 dimensional histogram with 0.2*0.2 mm bins). The main poles have been identified. The region of lower efficiency of the MCP can be identified on the left of the image, below the 113 pole.

Fig.4: Schematic view of the point projection of the atoms from the surface onto the detector. The green arrow indicates the angle between the 204 and 002 crystallographic directions, the red line highlights the effect of the compression factor, $\xi$, largely not to scale here, on the observed angle. The other parameters are defined in the text.

Fig.5: Compression factor (squares and gray line, left scale) and voltage (black line, right scale) as a function of the number of atoms. The corresponding depth has been indicated.

Fig.6: (a) 2D map showing the positions of the first 1500 ions arrived on the detector. (b) Schematic view of the successive terraces and the corresponding distances and angles.

Fig.7: Plot of 1-cos($\theta_n$) as a function of the number of the terrace and fitting curve. The error bars were estimated from measurements of the size of the successive terraces.

Fig.8: Measured radius of curvature as a function of the total voltage applied to the specimen. The radii were measured on 2D maps each containing 1500 atoms after the detection of; 6, 10, 15 and 25 millions atoms respectively.

Fig.9: 2D and 3D image (a, b, c) and Fourier transform (d) corresponding to all the atoms within the small volume shown in (a) and (b). The total volume considered here is 35x35x19 nm and only 1% of the atoms are represented, the small volume is 50x2x5 nm. In (c) the presence of multiple atomic plane families are clearly observable. On the diffraction pattern (d), the bordered dots correspond to the spot that can be observed. The direction of observation of the diffraction pattern is 110. This diffraction pattern has been obtained using all the best reconstruction parameters.

Fig.10: Measured value of $\theta_{[0\,0\,2]-[\bar{1}\,\bar{1}\,3]}$ as a function of the detection efficiency; the dashed-dotted line is included as a guide for the eye. The value of the crystallographic angle (25.24) has been highlighted.

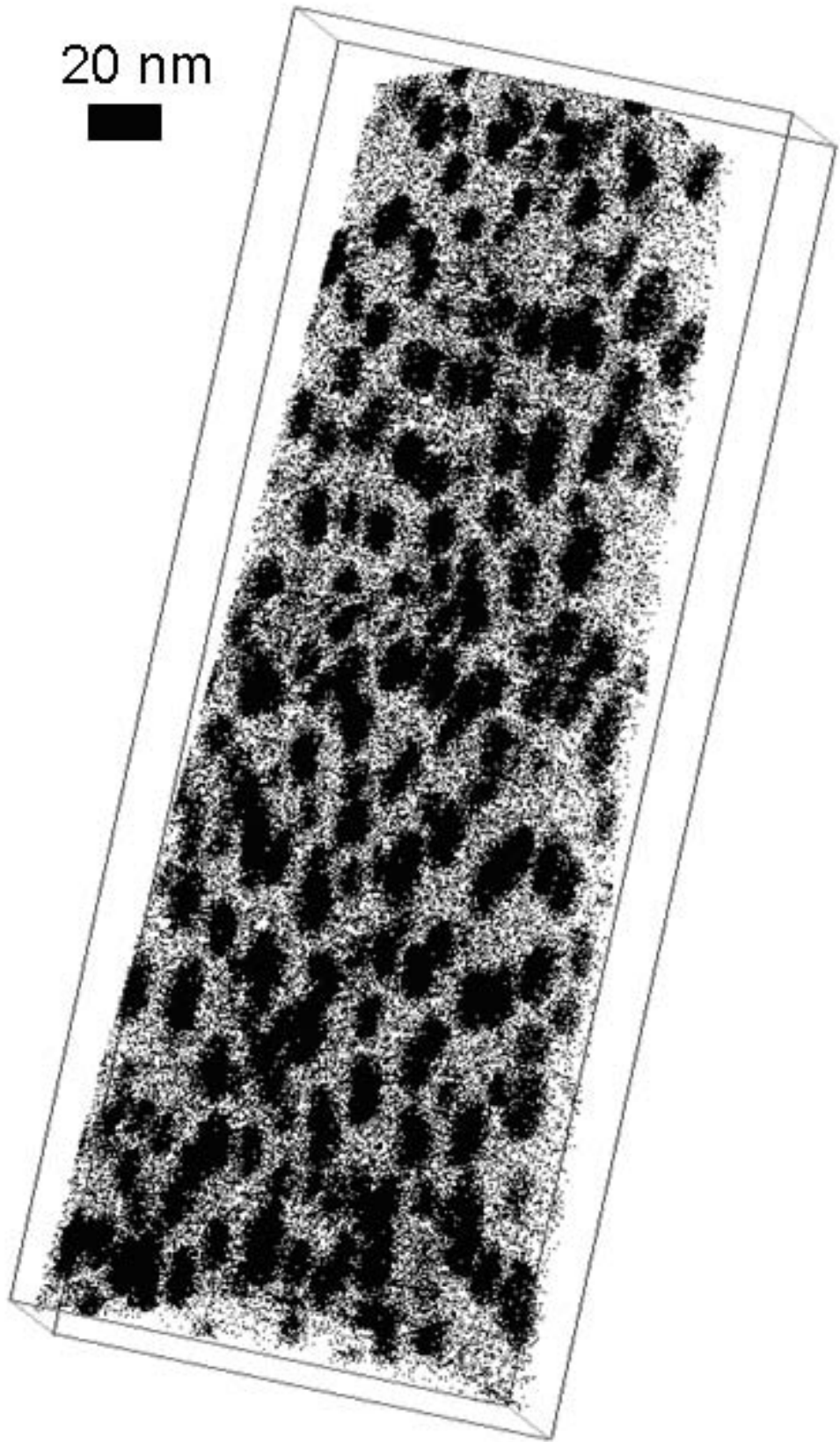

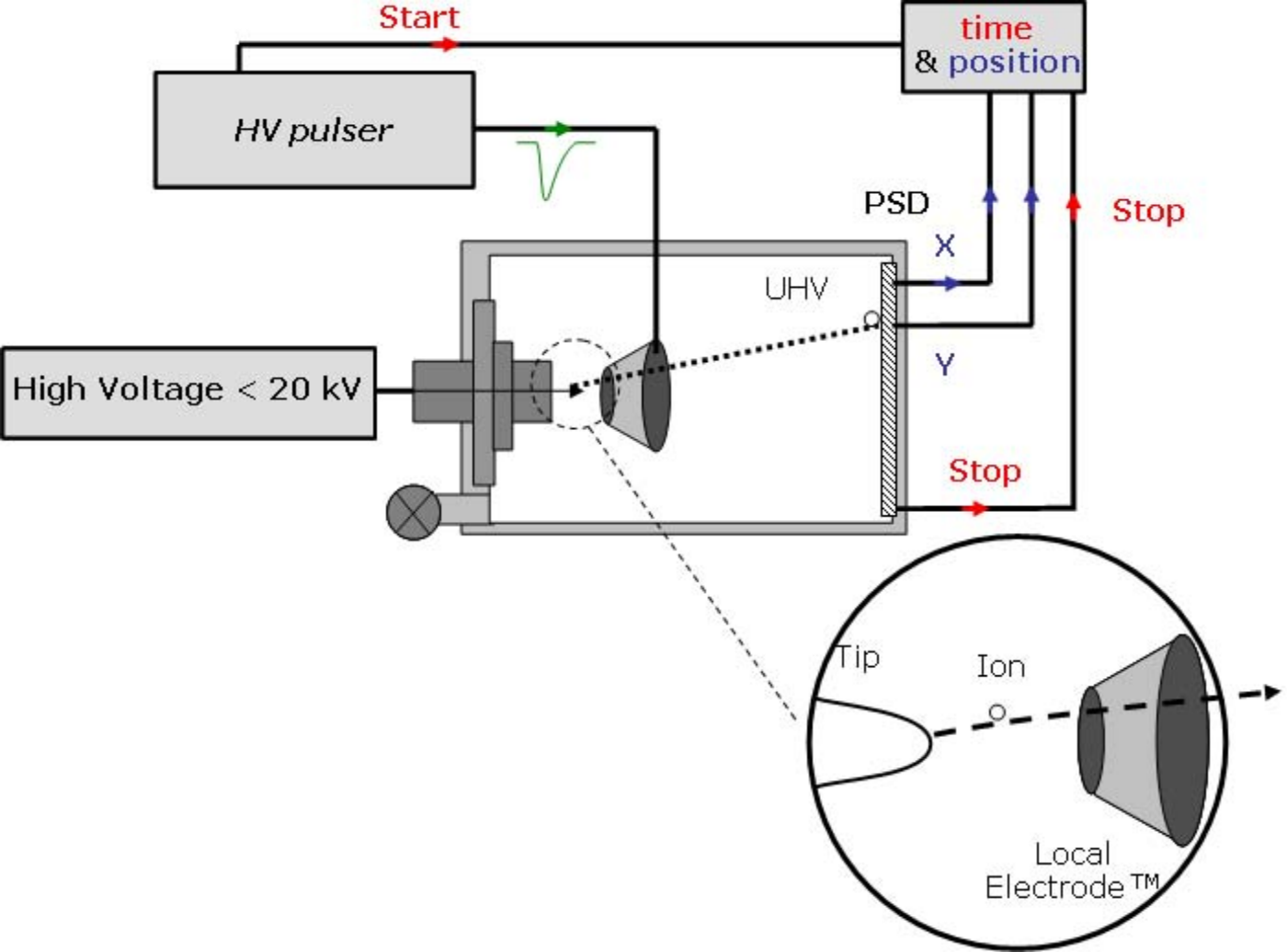

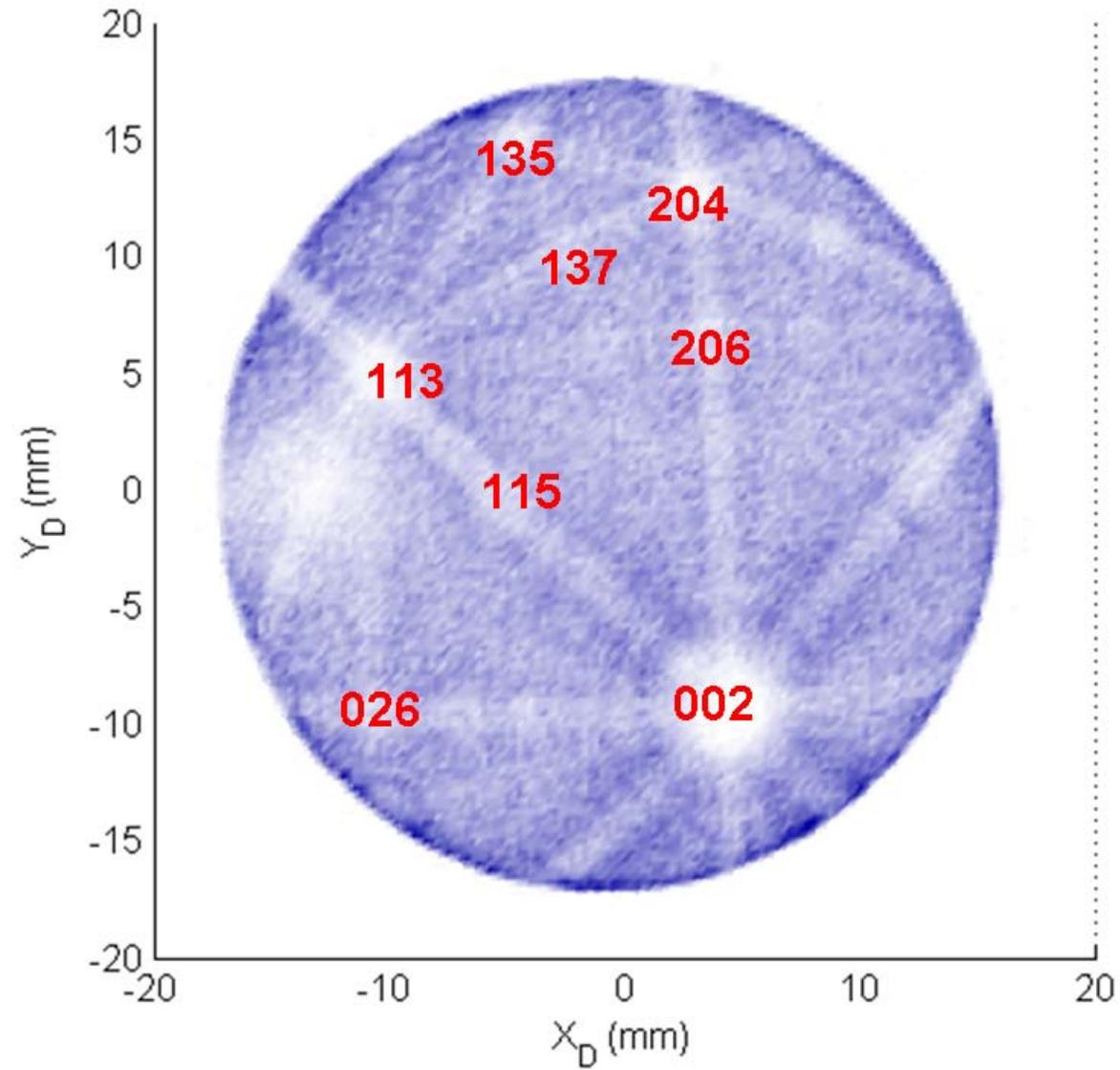

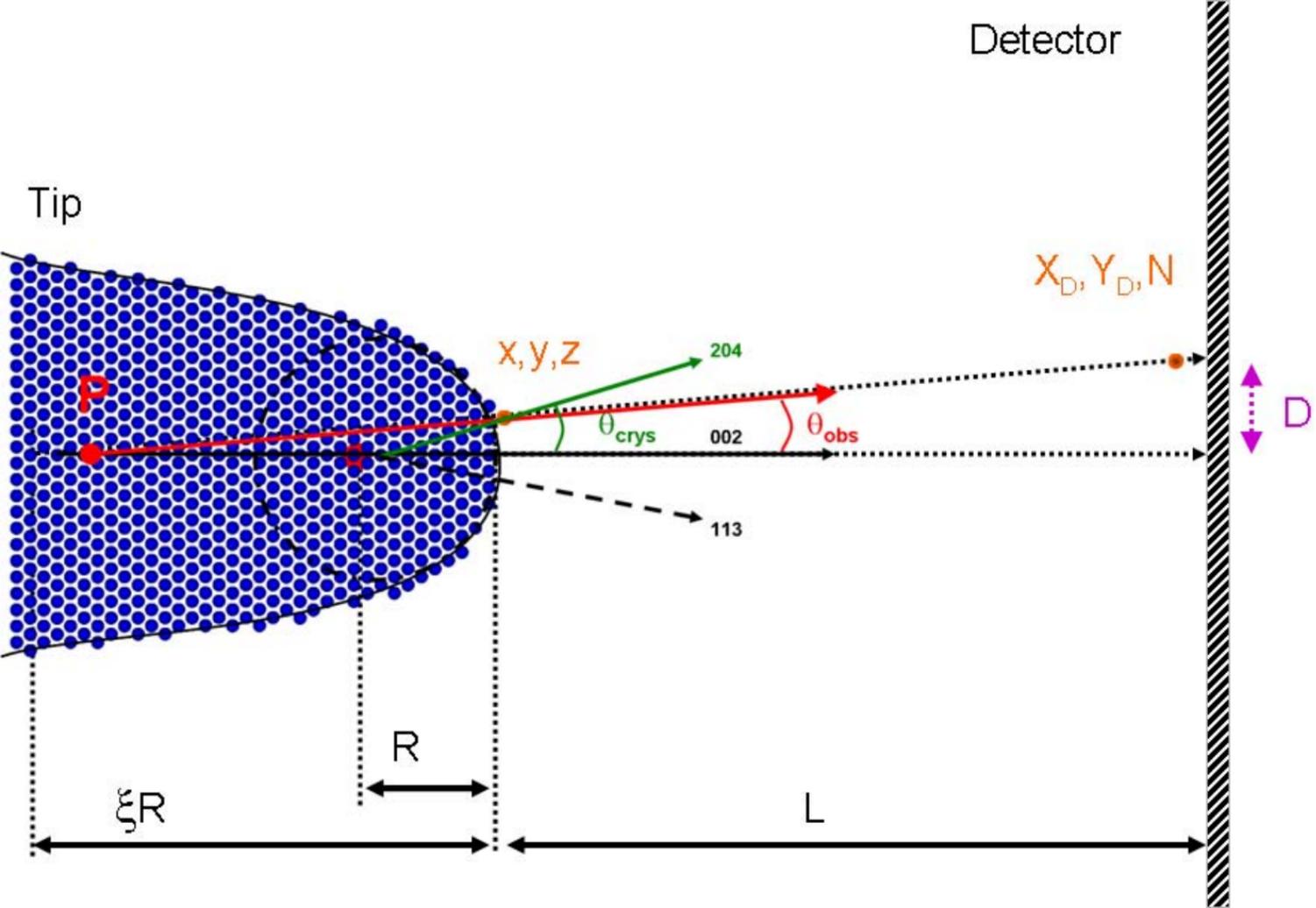

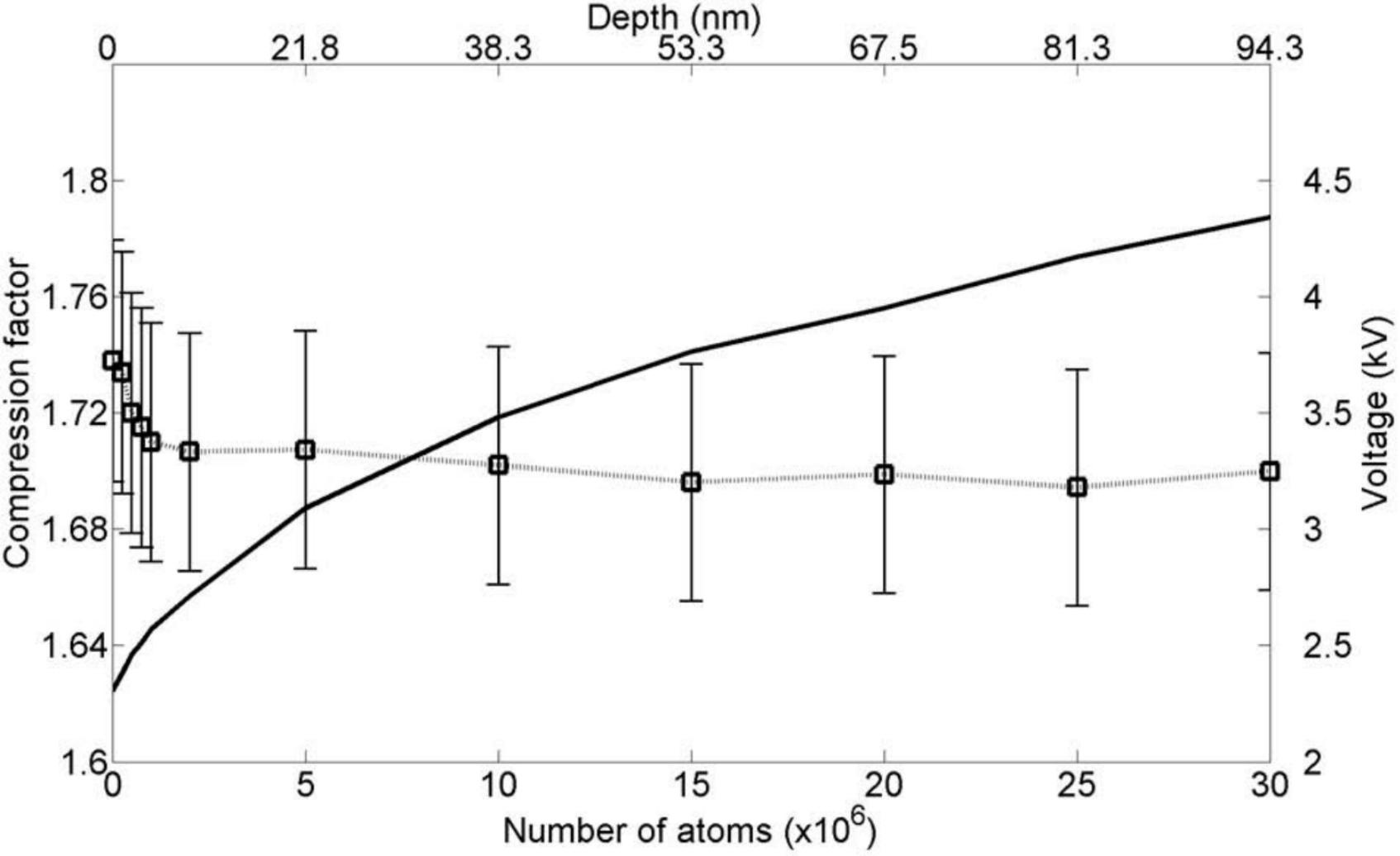

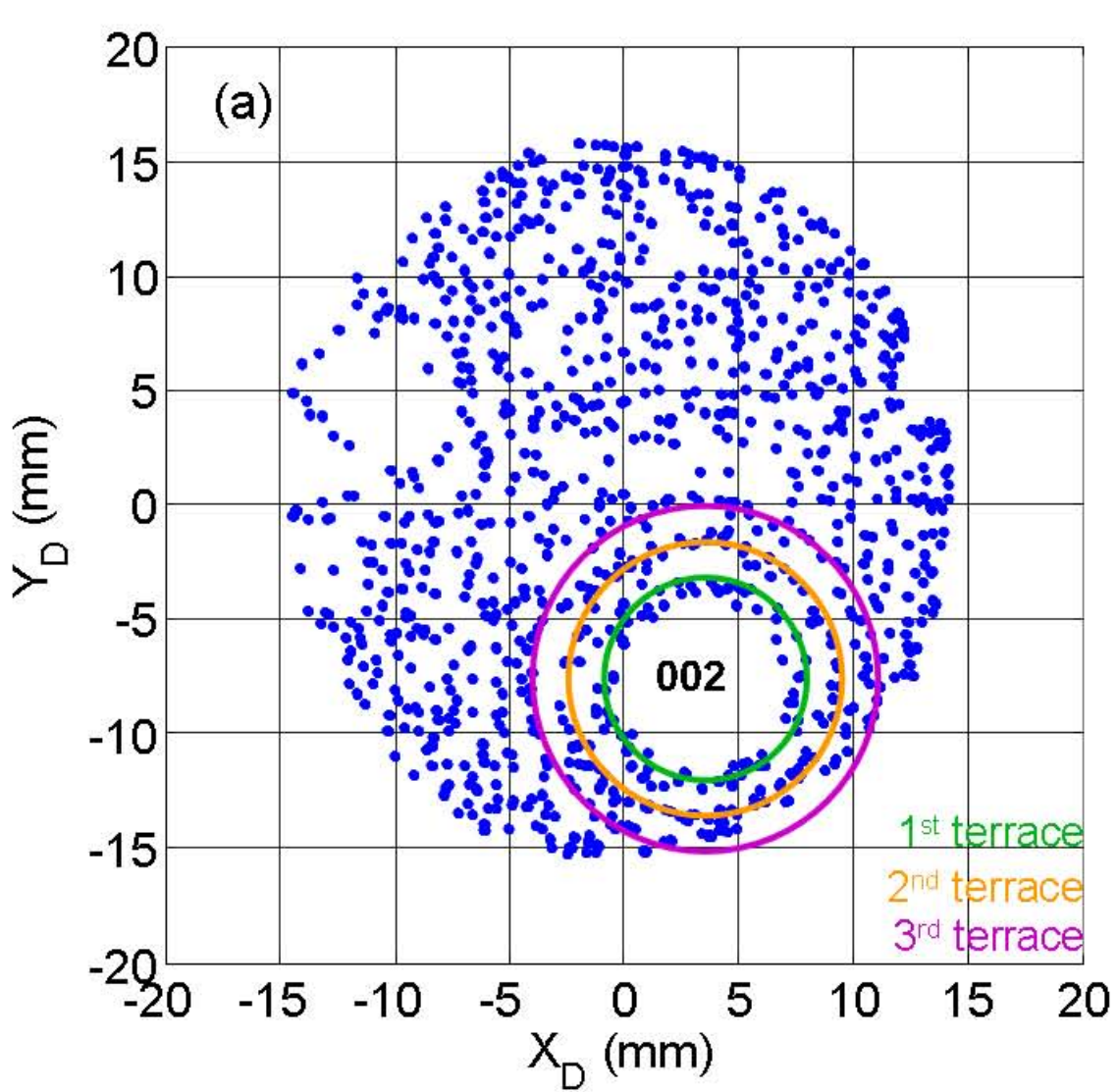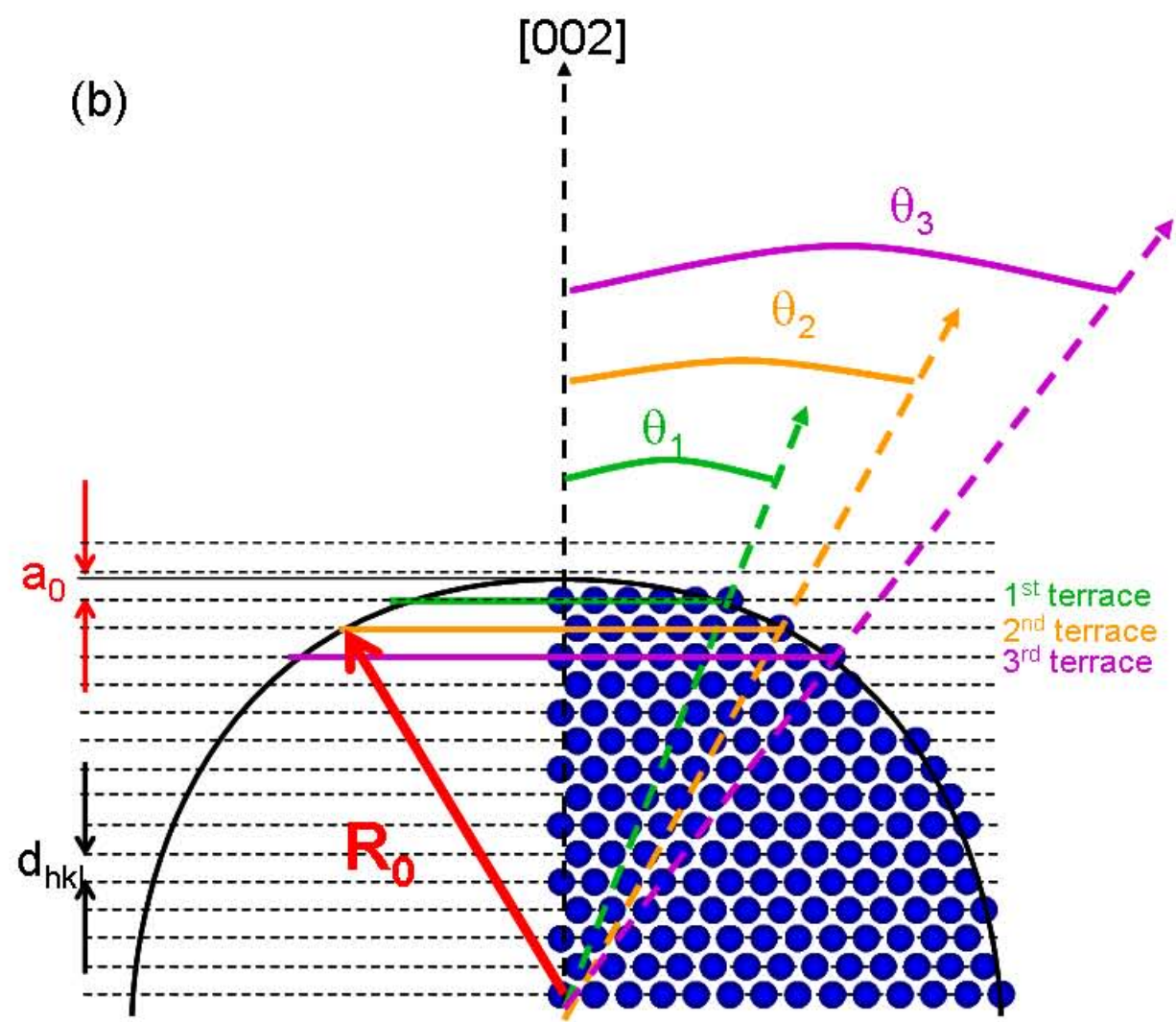

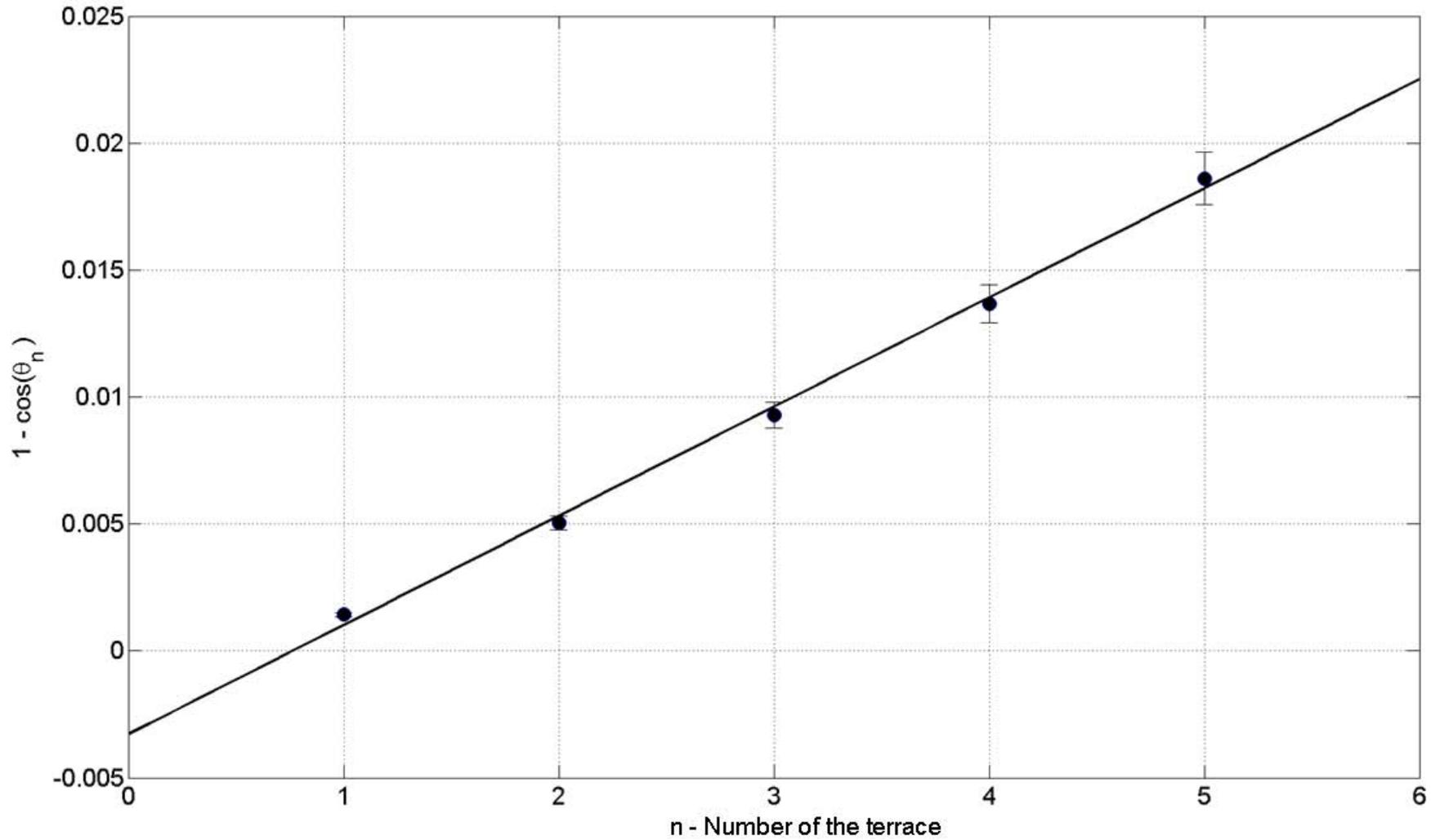

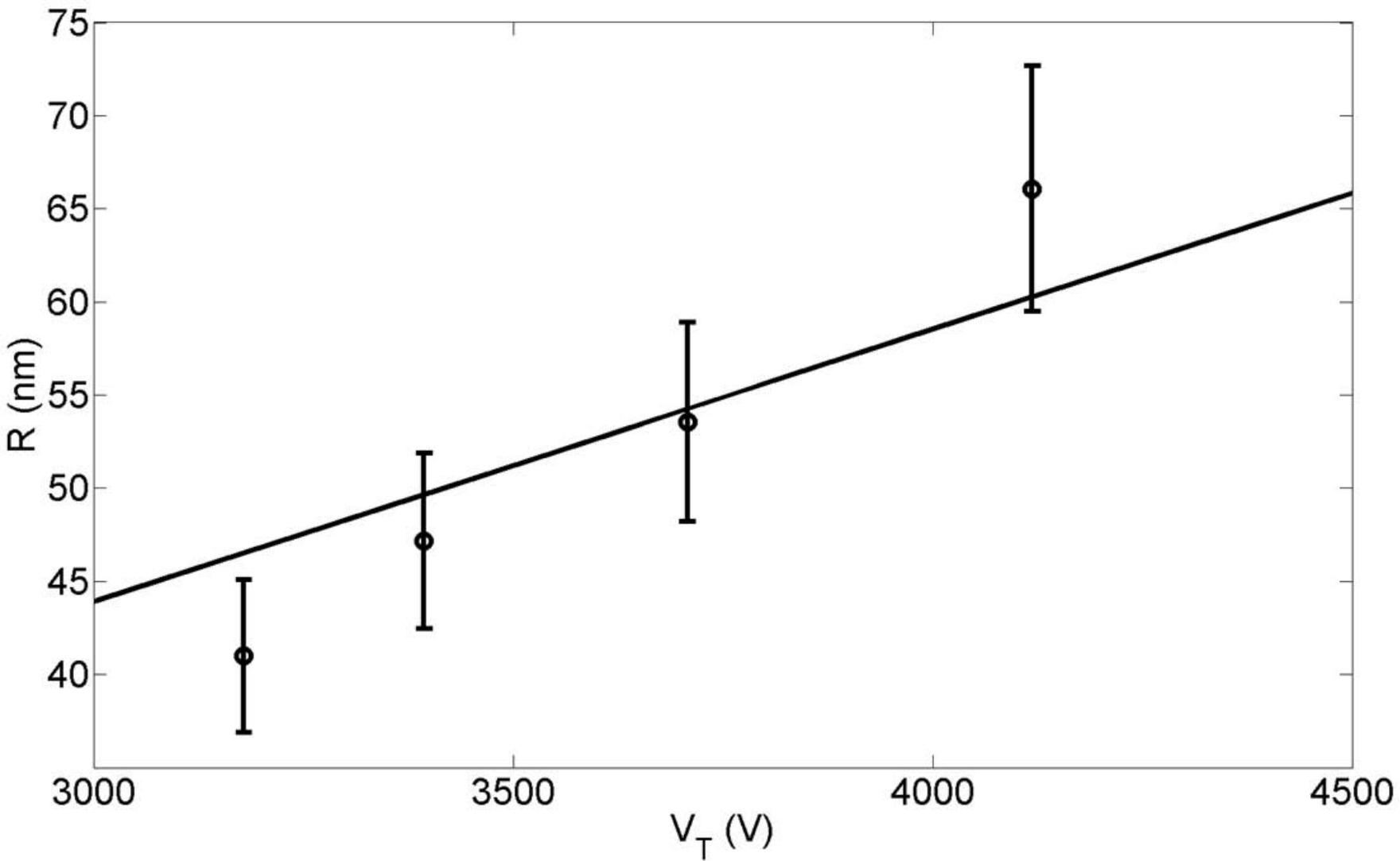

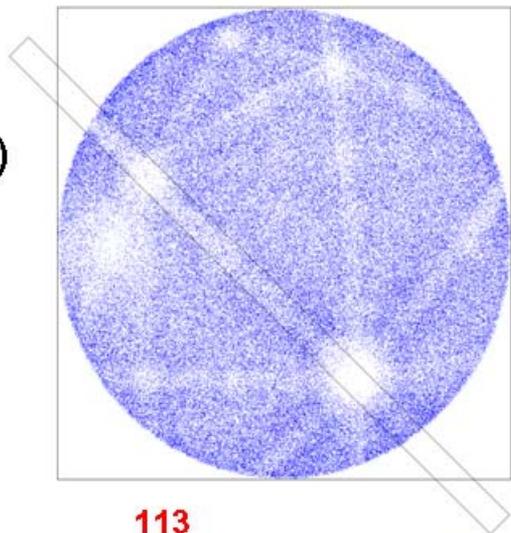
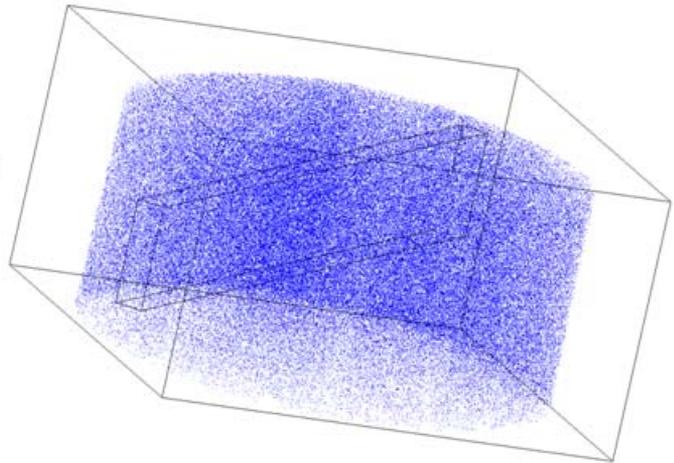

**113**                  **002**

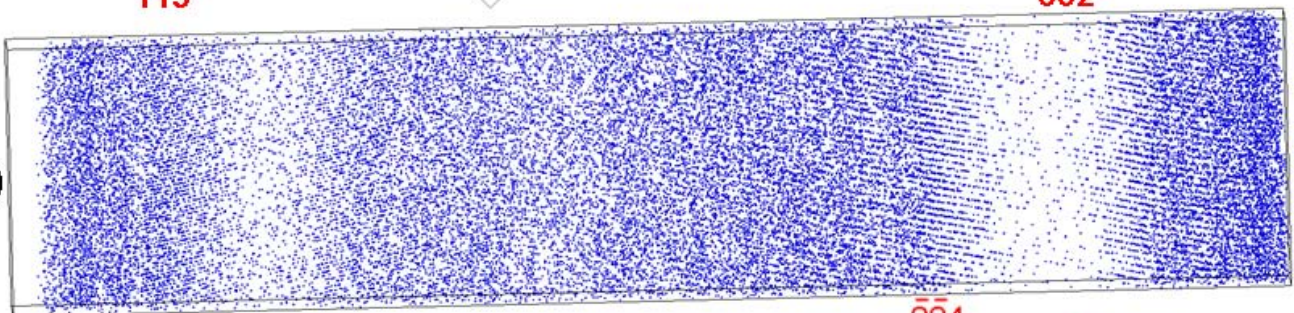
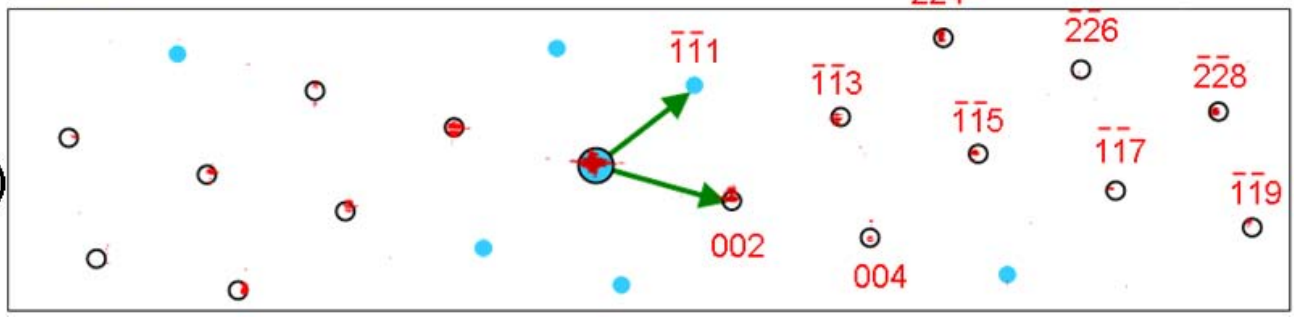

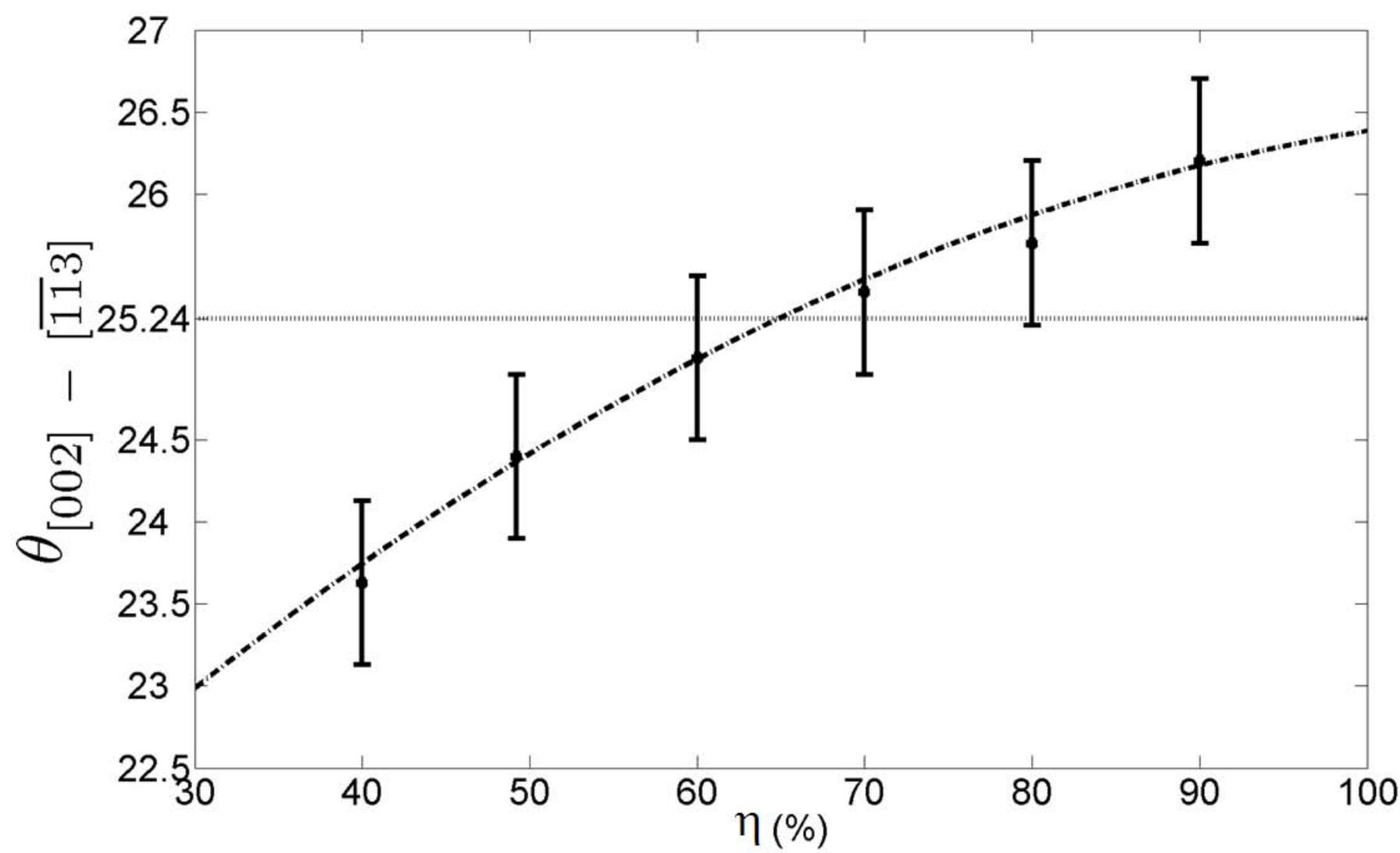

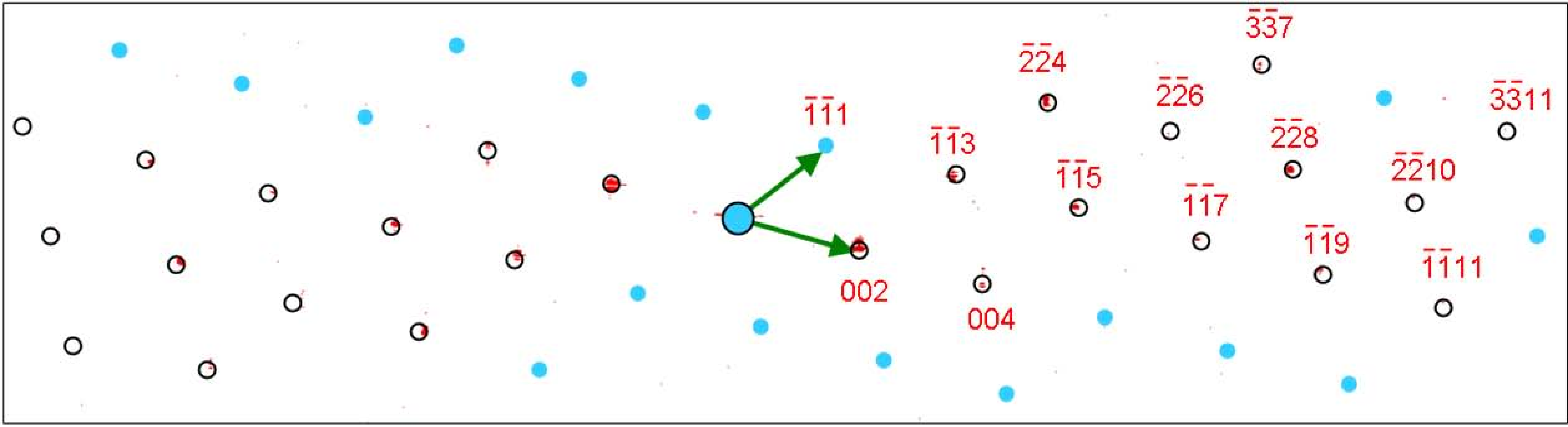